\documentstyle[twoside,fleqn,jhuwkshp,psfig]{article}
\newcommand{\etal}{{\it et al.} }

\newcommand{\xmm}{{\it XMM-Newton} }
\newcommand{\chandra}{{\it Chandra} }

\begin{document}

\title{Emission Line Diagnostics of Accretion Flows in Weak-Line Radio Galaxies}
 
\author{Karen T. Lewis \address{Department of Astronomy and Astrophysics,
The Pennsylvania State University, 525 Davey Laboratory, University
Park, PA 16802, USA.}, Michael Eracleous $^{{\rm a}}$, Rita
M. Sambruna
\address{Department of Physics and Astronomy, George Mason University,
4400 University Dr., Fairfax, VA 22030, USA} }

\begin{abstract}

In recent surveys of Radio-loud AGN, a new sub-class of objects, known
as Weak-Line Radio Galaxies (WLRGs) has emerged. These radio galaxies
have only weak, low-ionization optical emission lines. In the X-ray
band, these objects are much fainter and have flatter spectra than
broad-line and narrow-line radio galaxies. In these respects, WLRGs
are reminiscent of Low Ionization Nuclear Emission Regions
(LINERs). We have begun a multi-wavelength study of WLRGs to better
understand their possible connection to LINERs and the structure of
the accretion flow in both these systems. Here, we present new optical
spectra of a sample of WLRGs. We find that 81\% of the objects have
optical emission-line properties that are similar to LINERs,
indicating that these two classes of AGN may be related. Future high
resolution X-ray observations of WLRGs will be critical in determining
the true nature of the accretion flow in these objects.

\end{abstract}

\maketitle

\section{Introduction}

In a survey of southern radio galaxies, Tadhunter \etal \cite{t98}
find that the luminosity of the optical emission lines is strongly
correlated with the radio luminosity. An exception are a group of
objects, dubbed Weak-Line Radio Galaxies (WLRG), in which the [O~III]
line is an order of magnitude fainter than in galaxies of comparable
radio luminosity and redshift. The low-ionization [O~II] and H$\beta$
lines strengths, however, are not as drastically reduced in WLRGs.
This observation {\it suggests} that the optical emission lines
properties of WLRGs may be quite similar to those seen in LINERs.

\begin{figure}[!th] 
\centerline{\psfig{file=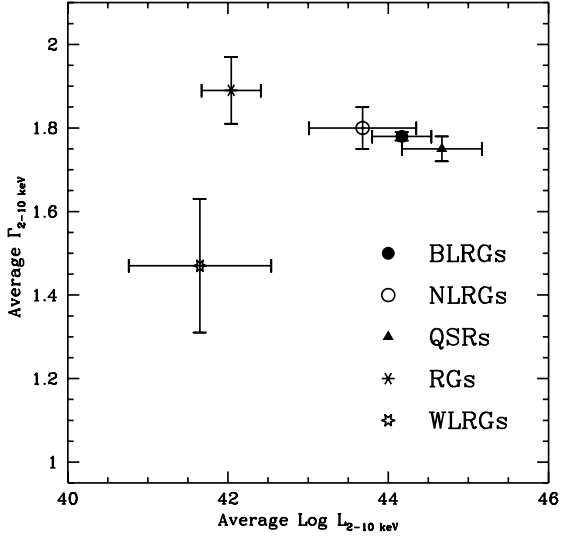,width=3.0in,height=3.0in}}
\caption{Average 2-10 keV photon index vs 2-10 keV 
intrinsic luminosity for the different classes of radio sources. WLRGs
have a significantly flatter average slope than the other
sources \cite{s99}.}
\label{wlrgfig}
\end{figure}

Even stronger evidence of a connection between WLRGs and accretion
powered LINERS is found in the X-ray and UV. In an analysis of ASCA
data Sambruna \etal \cite{s99} find that WLRGs are a distinct class of
object, set apart from the other radio-loud AGN\footnote{Throughout
this paper, radio loud AGN are those associated with powerful
double-lobed radio sources with L$_{rad} \ge 10^{25} W/\Omega$ at
5GHz.} by their low X-ray luminosity and flat X-ray spectrum
(Fig. \ref{wlrgfig}). Ho \cite{ho99} also finds that the ASCA spectra
of accretion-powered LINERs are somewhat harder than those of
radio-quiet AGN. However, it must be noted that the nucleus need not
dominate the emission from the galaxy in either WLRGs or LINERs
\cite{s00}. To determine the true X-ray characteristics of the central
engine, it is vitally important to obtain high spatial resolution
X-ray data, such as that obtained with \chandra and \xmm.

\begin{figure}[!bh] 
\centerline{\psfig{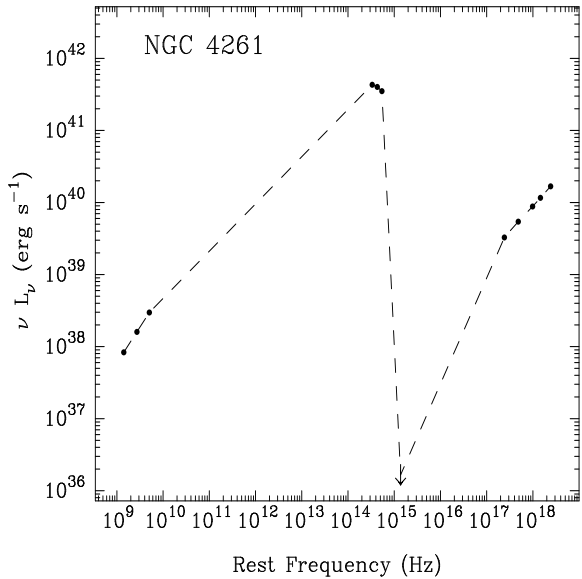}}   
\caption{The SED of the WLRG NGC 4261. Note the deficit of UV emission \cite{s99}.}
\label{wlrgsed}
\end{figure}

The spectral energy distribution (SED) of LINERs shows a significant
lack of UV emission. Although it is difficult to measure the SED of
WLRGs in the optical-UV band due to the significant absorption
column, a similar decrement is measured in at least one WLRG
(Fig. \ref{wlrgsed}). Thus there is mounting evidence that the optical
emission-line properties as well as the underlying ionizing continuum
are similar in WLRGs and LINERs.

Should a connection between the central engines of accretion-powered
LINERs and WLRGs be confirmed, this would place important constraints
on the structure of the accretion flow in these objects. Any proposed
model must be able to explain the behavior of both LINERs {\bf and}
WLRGs. In particular, the engine must be able to produce the
kpc-sized radio jets seen in WLRGs.

In this paper we present optical spectra of 11 WLRGs and demonstrate
that the emission line properties of WLRGs and LINERs are in fact
quite similar to each other. This work represents the {\it first step} in
our multi-wavelength program to better understand the central engines 
in WLRGs and the possible connection between LINERs and WLRGs.
\section{Optical Data}

\begin{figure}[!bh]
\centerline{\psfig{file=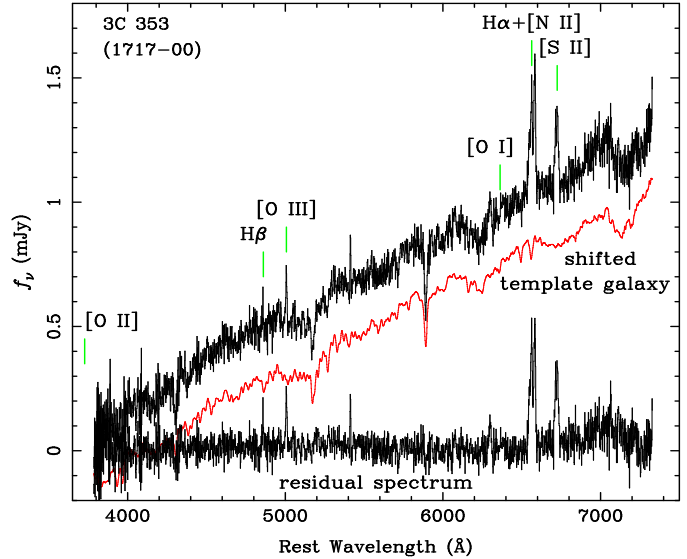,width=3.5in,height=2.9in}}
\caption{Example of galaxy template subtraction 
for 3C 353, which is an average spectrum in terms of initial line
strength, S/N, and resolution. Top: Original spectrum, Middle:
Template, Bottom: Residuals, from which the line fluxes are
measured. Important diagnostic emission lines are labeled.}
\label{1717}
\end{figure}

\begin{figure*} 
\centerline{\psfig{file=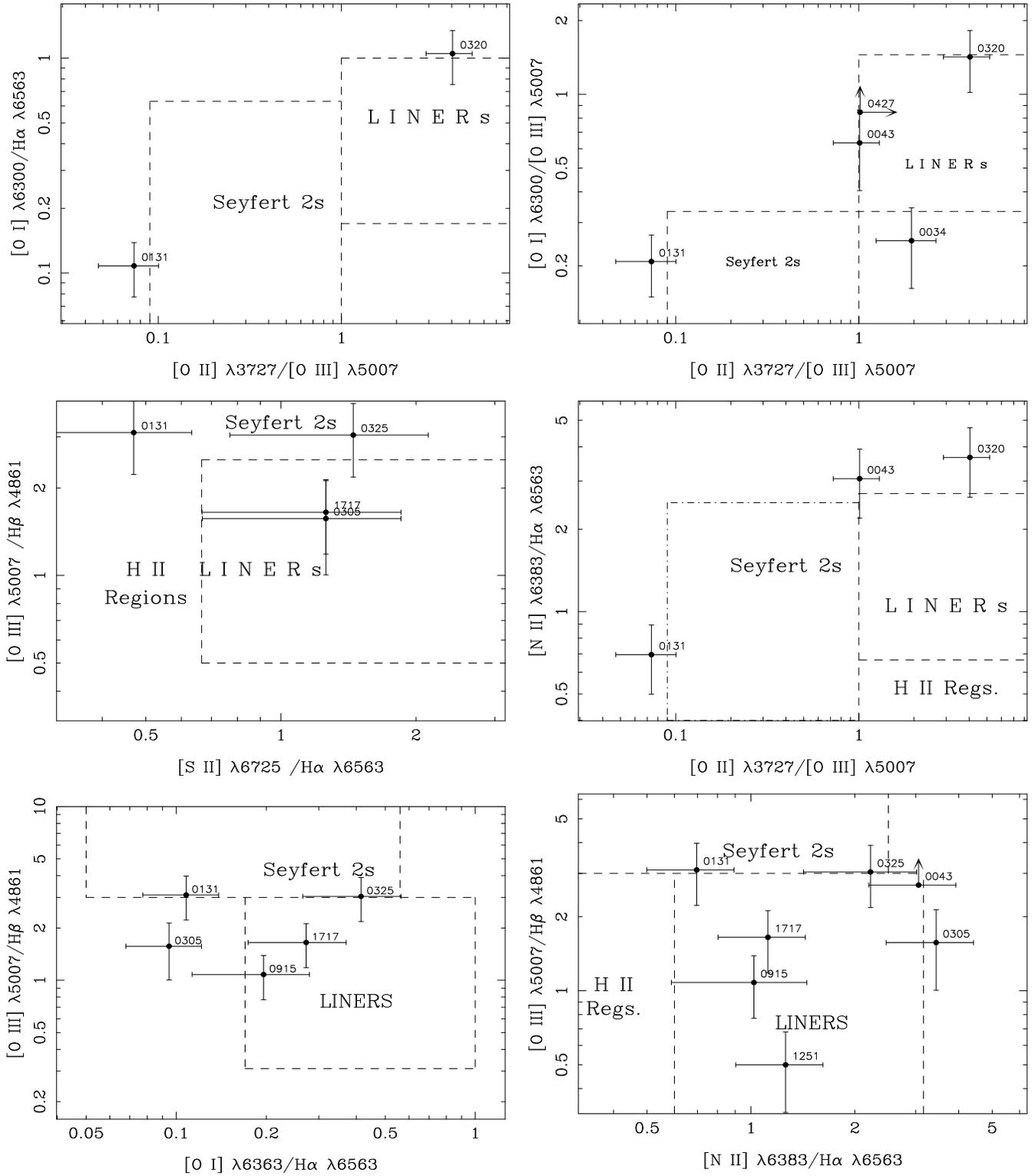,width=7.0in,height=8.0in}}   
\caption{Diagnostic diagrams for the 11 WLRGs. The regions generally 
occupied by LINERs, Seyferts, and H~II regions are designated. See
\cite{bpt,vo87} for definitions of the classification scheme.}
\label{diag}
\end{figure*}

The data set used here consists of spectra of 11 WLRGs, as listed in
Tab. 1. The spectra were obtained at KPNO 2.1m, CTIO 1.5m and the MDM
Observatory 2.4m telescopes. In order to measure the strengths of the
emission lines, the starlight from the host galaxy must be
subtracted. This was done by fitting a template elliptical galaxy to
the spectrum (adjusting the redshift and Galactic reddening to match
those of the object) and then subtracting the fit from the object
spectrum. An example of the template subtraction is shown in
Fig. \ref{1717}. Some residual continuum remains, which was subtracted
in the neighborhood of each emission line. Each emission line was then
fitted with a Gaussian profile. Finally, the fluxes measured from the
model profiles were used to place WLRGs in several classification
schemes for emission line galaxies, as shown in Fig. \ref{diag}
\cite{bpt,vo87}. These diagnostics have been developed {\it observationally}
over the past 20 years, as they cleary separate different classes
of emission line galaxies.

\section{Results}

All of the objects are classified in at least one of the six 2D
diagnostic diagrams (Fig. \ref{diag}) with the exception of
OH-342. We divide the objects into two main groups:

{\it Pure LINERs}: These objects fall within the LINER region
in all diagrams for which the necessary lines were measured. 

{\it Composite LINERs}: While showing evidence of LINER-like emission
in some diagrams, several objects are also classified as a Seyfert
and/or H~II region in others. Other objects consistently straddle the
boundary between LINERs and Seyferts or H~II regions. All of these
objects are classified as composites.

Most of the WLRGs (81$\%$) in our sample are classified as either a
pure ($36\%$) or composite (45$\%$) LINER (Tab. 1). NGC 612 is
classified as a Seyfert and OH-342, not shown, is likely to be a Seyfert
based upon its large [O~III]/H$\beta$ ratio.

\begin{table} 
\caption{Weak-Line Radio Galaxies in the Current Sample}
\label{wlrglist}
\vspace{12pt}
\begin{tabular}{llcc}
IAU     &                          &        & X-ray \\       
Name    & Other Names *            & $z$    & Obs.  \\
\cline{1-4}
0034-01 & 3C 15 $^{C}$             & 0.073  &       \\      
0043-42 &       $^{C}$             & 0.117  &       \\    
0131-36 & NGC 612 $^{S}$           & 0.030  &  ASCA \\  
0305+03 & 3C 78, NGC 1218 $^{C}$   & 0.029  &  SAX  \\
0320-37 & For A, NGC 1316 $^{C}$   & 0.006  &  ASCA \\
0325+02 & 3C 88 $^{C}$             & 0.031  &  SAX  \\
0427-53 &       $^{L}$             & 0.040  &       \\
0625-35 & OH-342 $^{S}$            & 0.055  &       \\
0915-11 & Hyd A, 3C 218 $^{L}$     & 0.054  &  Chandra \\
1251-12 & 3C 278 $^{L}$            & 0.016  &       \\
1717-00 & 3C 353 $^{L}$            & 0.030  &  ASCA \\
        &                          &        &       \\
\end{tabular}
\footnotesize{* L=Liner, S=Seyfert, C=Composite}
\end{table}

\section{Conclusions, Discussion, and Future Work}

We have shown that the optical emission line properties of most WLRGs
{\bf are} similar to LINERs. Several objects in our sample are
classified as composites due to the large error bars on the
measurements. Better optical spectra of some objects are needed to
solidify the classification. We have obtained additional spectra,
including new objects, which we are currently reducing and
analyzing. High spatial resolution spectroscopy from the ground or
with HST will also be helpful in this respect.

However, it is likely that WLRG, like LINERs are a diverse group of
objects. The central engine need not dominate the energetics of the
galaxy and other excitation sources such as a hot stars, shocks, or
cooling flows may play an important role in the formation of the
emission lines. In Hyd A, for example, UV observations reveal a
circumnuclear ring of star formation \cite{M95,M97}.

To determine whether WLRGs and LINERs are physically similar, it is
necessary to study the SEDs of WLRGs. In the UV and X-ray regimes this
is quite challenging. There appears to be a deficit of UV photons in
WLRGs (Fig. \ref{wlrgsed}), but the high absorption column (NH $\ge
10^{23} \rm cm^{2}$ in some cases) makes a UV flux measurement, or even a
useful upper limit, virtually impossible to obtain. Until recently,
the nuclear region of WLRGs and LINERs have not been spatially resolved in
X-rays, making it impossible to truly determine the X-ray portion of
the SED. A recent \chandra observation of Hydra A \cite{s00} has
demonstrated that while there can be several sources of x-ray emission
in WLRGs, the nucleus can be successfully isolated from the host
galaxy. Considering the difficulty of the UV observations, studies of
WLRGs in the X-ray regime are the best hope for constraining their SED.

The results of this work strongly suggest that LINERs and WLRGs may
be intimately related. More detailed studies in the UV
and particularly X-ray will solidify this relationship and lead to an
improved understanding of the structure of the accretion flow in both
LINERs and WLRGs.

\end{document}